# SDS : Une infrastructure d'installation de logiciels libres pour des organisations multi-sites.

**Laurent CHARLES, Manuel VACELET, Mohamed CHAARI, Miguel SANTANA**

*STMicroelectronics*
*850, rue Jean Monnet*
*F-38926 CROLLES CEDEX*
*laurent.charles@st.com*

*RÉSUMÉ. Les développements logiciels sur les systèmes UNIX font de plus en plus appel aux logiciels libres. Nous proposons une solution de déploiement et de contrôle de ces logiciels libres au sein d'une grande organisation. Nous nous attachons particulièrement à résoudre les problèmes liés au déploiement multi-sites ainsi qu'à la gestion de configuration de ces déploiements. L'originalité de notre approche repose sur sa capacité à être mise en œuvre et contrôlée par les utilisateurs plutôt que par les administrateurs, sans nécessiter d'expertise particulière, et par les possibilités de déploiement dans des environnements hétérogènes.*

*ABSTRACT. Free and open source software is more and more used for software developments on UNIX systems. We are proposing a solution to control the deployment of free software in the context of a large corporation, focusing on multi-site deployment and configuration management. The originality of our approach rests on its ability to be implemented and controlled by users rather than administrators, without requiring any particular expertise, and on its facility to be deployed in heterogeneous environments.*

*MOTS-CLÉS : Logiciel libre, Gestion de configuration, Déploiement logiciel, UNIX, Linux*

**1.     Introduction**

Les développements logiciels sur les plateformes UNIX utilisent de plus en plus les logiciels libres, en particulier des outils de développement. Dans de grandes organisations telles que les sociétés multinationales, l'usage et le déploiement de ces logiciels posent des problèmes de contrôle et de gestion de configuration.

SDS (*Software Delivery System*) propose une infrastructure permettant l'installation de logiciels libres par un processus automatique, pour chaque site de la société, chaque équipe, ou chaque projet. SDS garantit la simplicité, le contrôle et l'unicité du processus d'installation. SDS offre également un environnement collaboratif permettant à chaque personne ou équipe de contribuer à l'enrichissement du dépôt commun de logiciels.

Cet article présente le contexte qui a initié le projet SDS au sein de STMicroelectronics, et détaille les problèmes rencontrés au sein de cette société multinationale. Nous présentons ensuite le positionnement de SDS par rapport à quelques solutions existantes, principalement en regard des distributions Linux. Nous détaillerons enfin certains éléments techniques apportés par SDS, avant de conclure par la présentation des résultats actuels du projet.

**2.     Motivations du projet**

STMicroelectronics est une société dans laquelle il existe de très nombreux développements logiciels. Le nombre de ces projets logiciels est en croissance forte. Ces développements impliquent de nombreuses équipes, souvent situées sur des lieux géographiques différents.

La plupart des équipes travaillant sur des systèmes UNIX utilisent des logiciels libres pour leurs développements. On peut citer des outils comme gcc ou perl.

Ces équipes de développement ont souvent leurs propres versions de ces outils, installées dans les espaces des projets, ou plus rarement dans des espaces centralisés, pour un site ou une organisation.

De fréquents problèmes, dus à un contrôle insuffisant des outils de développements utilisés, ont été reportés lors des phases d'intégration par des équipes multi-sites, ou appartenant à des organisations différentes.

La charge de travail pour maintenir un environnement de développement fonctionnel est lourde pour les organisations. Installer et configurer ces logiciels requiert une expertise que n'ont pas forcément les équipes de développement.

La solution à ces problèmes passe par la possibilité pour les équipes de contrôler les logiciels utilisés, c'est à dire par une connaissance stricte, et une garantie de stabilité, des versions de ces logiciels.

Il est également important de proposer un système d'installation de logiciels qui soit simple et ne demande par une expertise que les développeurs ou les administrateurs système n'ont pas.

## 3. Travaux similaires

Dans le contexte du déploiement de logiciels libres, la première approche consiste à se tourner vers les mécanismes de distributions utilisés pour les systèmes Linux. Ces techniques sont séduisantes en regard des exigences annoncées, mais présentent certaines difficultés dans le cadre d'une mise en oeuvre globale au sein d'une grande entreprise.

L'hétérogénéité des environnements constitue un premier obstacle. Tous les projets de développement logiciels ne disposent pas en effet des mêmes environnements de travail, des mêmes systèmes d'exploitation. Notre objectif est d'offrir une solution pour tout système de type UNIX[1].

L'usage de logiciels libres doit s'intégrer dans les environnements habituels des développeurs. Le système, les outils utilisés, les plateformes et stations de travail, voire les habitudes des développeurs, ne doivent pas être modifiées.

Les organisations doivent en outre pouvoir rendre disponibles simultanément des versions différentes des logiciels, afin de répondre aux exigences de certains projets. Il est fréquent que deux projets utilisent des versions différentes d'un même outil, y compris au sein de la même organisation. Les projets de maintenance de produits anciens, ou les projets ayant des contraintes d'intégration dépendantes de différentes versions d'outils, sont dans ce cas. L'approche classique, qui propose une installation contrôlée dans un répertoire */usr*, n'est pas satisfaisante, car elle ne propose qu'une version donnée des outils.

La stabilité des versions d'un ensemble d'outils utilisés par un projet doit être assurée. Sauf exception, les outils de développement ne doivent pas être mis à jour pour suivre les dernières versions disponibles. Par exemple, la décision de changer de version de compilateur est une décision qui doit être sous complet contrôle du projet, et non sous contrôle d'administrateurs systèmes qui ne sont pas nécessairement au fait des contraintes du projet.

---

[1]. Nous choisissons d'écarter les systèmes Windows. Ces systèmes ont une approche très différente de l'organisation des outils de développement, et utilisent de façon beaucoup plus marginale les logiciels libres.

Il est indispensable d'avoir la possibilité de choisir de façon précise les versions des outils installés. S'il peut être acceptable pour un projet qui débute d'utiliser la dernière version disponible, un projet plus ancien a bien souvent des contraintes de gestion de configuration qui lui imposent une version donnée de certains outils. Ce point est particulièrement sensible pour résoudre les problèmes d'intégration. C'est la seule solution pour que deux équipes partenaires aient la possibilité de disposer d'outils identiques si elles sont situées sur des sites ou des réseaux différents.

Dans notre contexte, tout utilisateur (au sens UNIX) doit avoir la possibilité d'installer un ensemble de logiciels. Il ne doit pas être nécessaire de disposer des privilèges d'administrateur (*root*).

Nous sommes convaincus que l'appropriation et le succès d'une solution comme SDS passe par une grande simplicité de mise en œuvre. Notre objectif est de proposer une infrastructure et des outils disponibles pour l'ensemble de la société, que les projets ou les organisations privilégieront à une installation propre à chacun. La simplicité d'installation est impérative, notamment lors des phases critiques de première mise en oeuvre, où l'installation de logiciels est fortement dépendante du système hôte.

A notre connaissance, aucune distribution Linux, ou approche similaire associée à d'autres systèmes d'exploitation, ne permet de répondre telle quelle à nos exigences. Nous retenons néanmoins quelques approches dont le projet SDS s'inspire fortement.

- **Debian GNU/Linux** [DEBIAN]

Debian GNU/Linux constitue la principale source d'inspiration de l'approche SDS. Cette distribution est très intéressante quant à la gestion des paquets logiciels. Elle bénéficie également d'efforts de portage sur des architectures non Intel, voire des systèmes dont le noyau n'est pas un Linux. Elle a à ce titre des objectifs proches des nôtres. Elle n'intègre toutefois pas la possibilité de mettre en oeuvre de multiples installations facilement utilisables, et son initialisation reste assez fastidieuse.

- **Gentoo/Linux** [GENTOO]

Cette distribution fait le choix de proposer l'installation de logiciels à partir de la recompilation des sources de façon fortement paramétrable. Ceci nous semble être le seul moyen pragmatique pour permettre les installations multiples en mode utilisateur, sans avoir à modifier le coeur de certaines applications.

- **Solarpack** [SOLPACK]

Cette approche développée par Sun Microsystems est une adaptation pour les systèmes Sun Solaris de l'approche NetBSD [NETBSD]. Solaris est un de nos systèmes cibles. Nous bénéficions, grâce à Solarpack, d'une expérience significative de portage de logiciels libres sous Solaris.

## 4. L'approche SDS

Compte tenu des exigences et contraintes exposées au chapitre précédent, nous avons décidé de proposer une nouvelle approche. SDS n'entre toutefois pas dans le cadre d'une démarche de recherche proprement dite. SDS cherche à être une réponse pragmatique et innovante aux problèmes énoncés.

Notre premier objectif est de fournir un système d'installation de logiciels simple et systématique. Un ou plusieurs dépôts contenant les logiciels disponibles est librement accessible pour l'ensemble des organisations et sites de la société. Différentes versions de chacun des logiciels fournis sont disponibles sur les dépôts.

Par conception, c'est toujours le client SDS qui est à l'initiative de l'installation et du choix de version des logiciels. L'installation via SDS ne se fait jamais à l'initiative d'une administration centrale déployant une version donnée d'un logiciel sur l'ensemble des sites de la société.

La mise à disposition de multiples versions d'un ensemble de logiciels se fait simplement en les installant dans des espaces séparés sur le site des utilisateurs

SDS propose une gestion automatique des dépendances entre les logiciels. Il est possible d'exprimer pour un logiciel, un ensemble d'autres logiciels ou bibliothèques sans lesquels le premier ne peut fonctionner.

Enfin, une infrastructure permettant un développement coopératif est à la disposition de tous les utilisateurs, pour alimenter la base commune de logiciels disponibles.

### 4.1. Scénario

Illustrons le fonctionnement de SDS par un exemple (Figure 1).

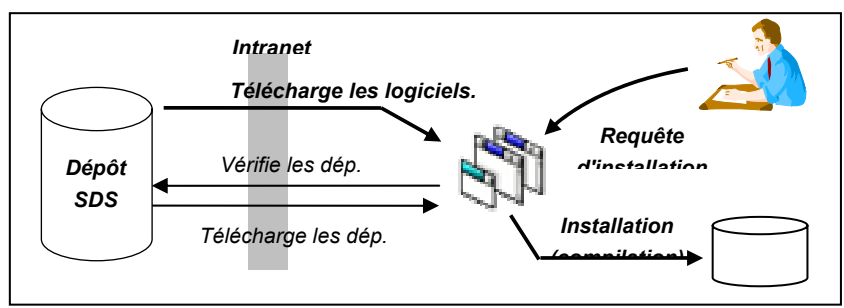

**Figure 1.** Principes généraux de SDS

Un utilisateur a besoin d'installer un ou plusieurs logiciels. Il a à sa disposition une application d'installation téléchargeable depuis le site intranet du projet SDS. Cette application lui permet de sélectionner un ensemble de logiciels à installer, en choisissant une version donnée ou la version la plus récente disponible.

L'application télécharge ensuite les logiciels sélectionnés. Un contrôle des dépendances de ces logiciels est effectué. S'il y a lieu, l'ensemble des dépendances est également téléchargé. Ce processus se répète récursivement tant qu'il existe des dépendances non résolues.

L'installation des logiciels s'effectue alors dans l'espace disque indiqué lors de la requête initiale. Cette installation est effectuée par compilation pour la quasi totalité des logiciels disponibles.

L'utilisateur a ainsi installé un ensemble de logiciels à l'aide d'une seule et unique commande. Il n'a eu besoin d'aucune expertise autre que de connaître les noms, voire les versions, des logiciels désirés. Tous les utilisateurs qui souhaiteraient installer ce même ensemble d'outils ont l'assurance d'utiliser le même processus d'installation.

*4.2. Architecture*

SDS est basé sur un système de gestion de paquets propre, avec une forte influence des systèmes de gestion de logiciels des distributions Debian GNU/Linux, Gentoo, et Solarpack. L'infrastructure SDS s'articule autour de deux composants : un système de gestion de paquets logiciels et une infrastructure web de gestion du dépôt des logiciels.

En pratique, il existe deux outils principaux (Figure 2) :

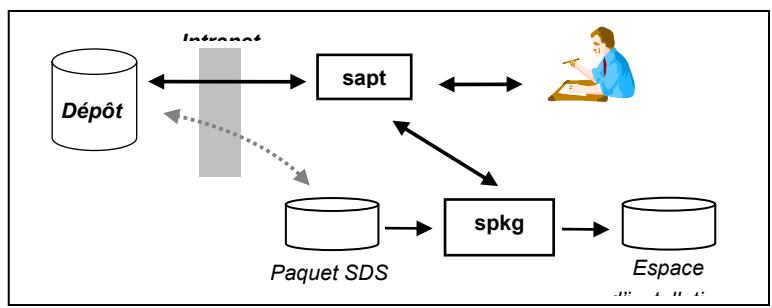

**Figure 2.**    Architecture de SDS

*spkg* : C'est l'outil de traitement du paquet SDS, le seul à avoir accès au cœur du paquet. Il est en charge de l'installation du logiciel en fonction des données présentes dans le paquet.

*sapt* : Cet outil contrôle l'approvisionnement en paquets lors d'une requête et gère les accès à l'infrastructure intranet. Il garantit à *spkg* toutes les conditions pour pouvoir installer complètement le paquet. C'est *sapt* qui gère en particulier les dépendances entre logiciels. Dans un usage normal, l'utilisateur de SDS n'a besoin que de *sapt*.

### *4.3. Paquets logiciels SDS*

SDS intègre une technologie de gestion de paquets logiciels à l'image de la plupart des distributions Linux. Le coeur du paquet est constitué principalement de quatre types d'informations :

Identification. Il s'agit d'un ensemble de fichiers identifiant précisément le logiciel : nom, version, description, licence, plateformes sur lesquelles il fonctionne.

*Upstream*. Il s'agit de la référence vers le logiciel empaqueté. Il se présente soit sous la forme du logiciel lui-même (souvent un *tarball* compressé) ou de l'URL permettant de télécharger le logiciel.

Dépendances. Nous disposons d'un langage simple permettant l'expression de règles de dépendances entre les logiciels. Pour pourvoir installer un logiciel, il est nécessaire que l'ensemble de ses dépendances soit installées. Il est possible d'exprimer des dépendances envers des logiciels et leurs versions.

Installation. Pour chaque logiciel est spécifié le processus d'installation suivant un canevas générique. Il s'agit d'un ensemble de programmes - des scripts pour des raisons de facilité de mise en oeuvre et de portabilité - qui seront exécutés lors de chacune des étapes de l'installation.

Le principe de notre système de paquets repose sur le postulat que tout logiciel peut s'installer en suivant un certain nombre d'étapes prédéfinies, telles que *extraction des sources*, *configuration*, *compilation*, *installation*, etc. En étendant ce principe à un logiciel empaqueté, nous définissons six étapes principales[2] :

- Extraction du paquet
- Test des dépendances
- Configuration
- Fabrication

---

[2]. Il existe en fait d'autres étapes, mais pour des raisons de simplification de notre propos, nous ne présentons que les six étapes majeures.

- Installation
- Enregistrement du logiciel

A chacune de ces étapes correspond un comportement par défaut. Si le logiciel nécessite une phase d'installation particulière, il suffit d'ajouter dans le paquet un programme effectuant cette opération. Ce programme doit simplement avoir un nom conventionnel en fonction de l'étape à laquelle cela correspond.

Lorsqu'un logiciel est installé, il est référencé dans une petite base de données présente dans le répertoire d'installation. La recherche de dépendances non résolues se fait en interrogeant ces données. Ainsi, un logiciel déjà installé n'est pas réinstallé, sauf si l'expression de la dépendance le rend obsolète.

Il est également possible pour chaque phase de définir une opération à effectuer avant (*pre-*) ou après (*post-*).

SDS considère a priori que les logiciels à installer respectent le "paradigme GNU" *configure/make/install*. Ainsi, pour tous ces logiciels, la mise en paquet est réduite au renseignement des informations d'identification du paquet et des dépendances. Pour tous les logiciels qui ne s'installent pas de cette manière, il suffit d'écrire certains des scripts d'installation afin de définir le comportement attendu.

Ainsi, par exemple, le paquet SDS de *gcc* a la structure de la figure 3.

```
gcc
   identification            depends
      NAME                      depends
      VERSION                sds
      LICENSE                   configure
      PLATFORM                  build
      MAINTAINER                post-install
   pkg
      upstream.url
```

**Figure 3.**   Paquet *gcc*

Les informations d'identification ainsi que les dépendances sont documentées dans les répertoires *ad hoc*. Le logiciel *gcc* lui même n'est pas inclus dans le paquet dans notre exemple, seule l'URL d'origine est renseignée. Lors de l'installation, les opérations de configuration (*configure*) et de compilation (*build*) sont différentes du processus par défaut. Une opération supplémentaire est également à effectuer une fois la phase d'installation terminée (*post-install*).

Cette structure de paquet permet également d'empaqueter tout logiciel proposant un processus d'installation différent. Il est nécessaire alors d'écrire les scripts modifiant le comportement par défaut. On peut illustrer ce point avec deux types de paquets particuliers : les paquets binaires, et les paquets virtuels.

Les paquets binaires incluent des logiciels sous forme binaire plutôt que sous forme de sources à recompiler. La fabrication d'un tel paquet se résume principalement à vérifier que l'installation se déroule sur une plate-forme compatible (script *pre-extract* par exemple), ne rien faire lors des phases de configuration et de compilation (scripts *configure* et *build* vides), et l'exécution d'un programme particulier avec des paramètres ad hoc, lors de la phase d'installation (script *install*).

Les paquets virtuels n'intègrent aucun logiciel, mais disposent d'un ensemble de dépendances. Ces paquets peuvent regrouper un ensemble d'outils sous un même nom, impliquant par exemple l'installation de dizaines de logiciels par le mécanisme des dépendances. Un projet peut disposer d'un paquet *gnu_projet* spécifiant l'ensemble des outils requis. Les équipes partenaires installent ce paquet et sont assurées de la cohérence de leurs outils de développement.

## 5.   Résultats

Le projet SDS a aujourd'hui deux ans d'existence, dont une année de déploiement effectif au sein de la société STMicroelectronics. Nous sommes en mesure de faire un premier bilan, d'un point de vue déploiement et d'un point de vue technique.

### *5.1.   Usage*

Au sein de STMicroelectronics, SDS est utilisé sur différents sites dans le monde. Certains sites ont fait le choix de fournir l'ensemble des logiciels libres utilisés par l'intermédiaire de SDS. La mise à disposition d'outils est sous la responsabilité des administrateurs systèmes ou de personnes identifiées.

Sur d'autres sites, principalement ceux qui ont un historique important, et pour lesquels il existe une base installée de logiciels libres, ce sont les projets qui font usage de SDS selon leurs besoins. Ils installent et contrôlent les outils dont ils ont besoin dans leurs propres espaces.

### *5.2.   Modes d'utilisation*

Nous constatons deux utilisations distinctes des logiciels libres proposés.

Les environnements utilisateurs. Dans ce cas, l'objectif des personnes qui utilisent SDS est d'installer un maximum de logiciels pour une communauté d'utilisateurs. Ces logiciels sont davantage des outils de productivité personnelle, tels que des environnements graphiques et suites bureautique, que des outils de développement. Ces utilisateurs cherchent à disposer des versions les plus récentes.

Les environnements de développement. L'objectif est dans ce cas de fournir des ensembles d'outils de développements stables et contrôlés pour les projets. Il n'y a pas de recherche de la mise à jour. C'est également ce mode de fonctionnement qui est utilisé par les équipes ayant des problèmes d'intégration Elles décident en commun de l'ensemble des outils à utiliser et installent chacun de leur côté des "clones" du même ensemble d'outils, dans les mêmes versions.

### 5.3. Système de gestion de paquets SDS

SDS utilise son propre système de gestion de paquets logiciels. Il est légitime de s'interroger sur la pertinence d'un tel choix.

Notre système de gestion de paquets est très simplifié par rapport aux autres approches. Nous n'avons pas l'ensemble des facilités habituellement disponibles. Mais dans notre contexte, nous n'avons pas trouvé de limitation importante. Aucun des logiciels disponibles par SDS n'a posé de problème d'adaptation. En revanche, la grande simplicité de notre approche permet de créer un paquet très rapidement, y compris pour un utilisateur novice. Dans le cas des logiciels GNU [GNU], il faut une dizaine de minutes pour intégrer un nouveau logiciel dans le dépôt.

Nous avons également gagné beaucoup de temps au niveau du développement et du portage. Les contraintes de portage auraient en particulier été très fortes vis à vis de nos objectifs multi plateformes. En quelques mois, nous avons pu avoir un système opérationnel.

### 5.4. Disponibilité

SDS est un logiciel libre, sous licence GPL. Il n'est toutefois pas diffusé aujourd'hui faute d'infrastructures d'hébergement publiques. SDS pourra être mis à disposition sur demande d'une organisation qui pourrait fournir ces infrastructures d'hébergement et de collaboration.

### 6.   Conclusion

Nous avons présenté dans cet article l'infrastructure SDS. Elle permet un déploiement multi-site automatisé et contrôlé de logiciels libres. SDS est utilisé maintenant depuis plus d'un an au sein de STMicroelectronics, sur plusieurs sites de la société.

Son coût de mise en œuvre et de maintenance est faible en regard de sa portée. Il n'y a pas de coût de licence, les infrastructures nécessaires sont simples et bon marché, son coût de développement est raisonnable.

Notre infrastructure est ouverte et libre. Nous bénéficions ainsi de retours et de contributions de nos utilisateurs, et pouvons leur apporter des réponses spécifiques.

SDS correspond bien à nos besoins et propose un bon compromis entre facilité d'utilisation et richesse fonctionnelle. Notre approche pragmatique apporte des réponses immédiates aux problèmes de gestion de configuration des équipes de développement logiciels.

## 7.    Perspectives

SDS ne satisfait bien entendu pas la totalité de nos besoins. Nous avons identifié un certain nombre d'axes de développement. Le produit SDS est en constante évolution.

Il est prévu à court terme de travailler sur l'amélioration de la gestion de dépendances. Nous voulons exprimer des relations de dépendances plus complètes que celles que nous gérons actuellement (arithmétique complexe et dépendances *depends*, *recommends*, *suggest*). Nous désirons à décorréler le logiciel lui-même de la fonction qu'il remplit (fonction *provide*), et souhaitons gérer la notion de dépendances « a posteriori » (*post-install*).

Nous avons également besoin d'améliorer les possibilités d'intégration des installations SDS avec l'environnement général. Cela passe par la possibilité de reconnaître de multiples installations SDS comme base pour la gestion des dépendances, par la possibilité d'intégrer des logiciels qui ne sont pas gérés par SDS dans la gestion des dépendances. Nous étudions également la possibilité de gérer les versions multiples des logiciels installés avec un outil comme Stow [STOW].

Nous aimerions enfin pouvoir confronter SDS avec d'autres besoins, d'autres approches, d'autres « marchés ». Cela passe sans doute par la possibilité de diffuser SDS, et la mise en place de collaborations externes à STMicroelectronics.

## 8.    Bibliographie

[DEBIAN]    Debian GNU/Linux, « Système d'exploitation libre », 'http://www.debian.org/'

[GENTOO]    Gentoo Linux, « Gentoo Linux, a special flavor of Linux that can be automatically optimized and customized for just about any application or need. », 'http://gentoo.org/'

[GNU] GNU, « Un système d'exploitation complet, semblable à Unix et qui soit un logiciel libre: le système GNU », 'http://www.gnu.org/'

[NETBSD]    NetBSD, « A free, secure, and highly portable UNIX-like Open Source operating system available for many platforms. », 'http://www.netbsd.org/'

[SOLPACK]    Solarpack, « A repository of 1400+ pre-compiled packages of free software for Solaris systems, a distribution of the NetBSD package collection tools (NPCT) for Solaris, the home of the Zounds package manager. » 'http://solarpack.sourceforge.net/'

[STOW] GNU/Stow, « A program for managing the installation of software packages, keeping them separate while making them appear to be installed in the same place. » 'http://www.gnu.org/software/stow/'